\documentclass[aps,superscriptaddress,showpacs,amsmath,amssymb,longbibliography,twocolumn,floatfix]{revtex4-1}
\usepackage{times}
\usepackage{setspace}
\usepackage{comment}
\usepackage[utf8]{inputenc}
\usepackage{graphicx}
\usepackage{color}
\usepackage{xcolor}
\usepackage{soul}
\usepackage{environ} 
\usepackage{float} 
\usepackage{mathtools}
\usepackage{natbib}
\usepackage[english]{babel}
\usepackage[pdftex, colorlinks=true, linkcolor=blue, citecolor=blue,urlcolor=blue]{hyperref}
\usepackage{microtype}

\newcommand{\en}{{\varepsilon_{n\textbf{k}}}}

\begin{document}
 
\title{Charge-carrier thermalization in bulk and monolayer CdTe from first principles}
\author{Dinesh Yadav}
\affiliation{Institute of Physics, University of Augsburg, 86135 Augsburg, Germany}
\affiliation{Okinawa Institute of Science and Technology Graduate University, Onna-son, Okinawa 904-0495, Japan}
\author{Fabian Pauly}
\affiliation{Institute of Physics, University of Augsburg, 86135 Augsburg, Germany}
\affiliation{Okinawa Institute of Science and Technology Graduate University, Onna-son, Okinawa 904-0495, Japan}
\author{Maxim Trushin}
\affiliation{Centre for Advanced 2D Materials, National University of Singapore, 6 Science Drive 2, Singapore 117546}

\date{\today} 

\begin{abstract}
While cadmium telluride (CdTe) thin films are being used in solar cell prototyping for decades, the recent advent of two-dimensional (2D) materials challenges the fundamental limit for thickness of conventional CdTe layers.
Here, we report our theoretical predictions on photocarrier dynamics in an ultimately thin (about 1 nm) CdTe slab. 
It corresponds to a layer that is just a single unit cell thick, when the bulk parent crystal in the zinc blende phase is cleaved along the [110] facet. 
Using an \textit{ab initio} method based on density functional theory (DFT) and the Boltzmann equation in the relaxation time approximation (RTA), we determine the thermalization time for charge carriers 
excited to a certain energy for instance through laser irradiation. 
Our calculations include contributions arising from all phonon branches in the first Brillouin zone (BZ), thus capturing all relevant inter- and intraband carrier transitions due to electron-phonon scattering.
We find that the photocarrier thermalization time is strongly reduced, by one order of magnitude for holes and by three orders of magnitude for electrons, once the CdTe crystal is thinned down from the bulk to a monolayer. 
Most surprisingly, the electron thermalization time becomes independent of the electron excess energy up to about 0.5~eV, when counted from the conduction band minimum (CBM).
We relate this peculiar behavior to the degenerate and nearly parabolic lowest conduction band that yields a constant density of states (DOS) in the 2D limit. 
Our findings may be useful for designing novel CdTe-based optoelectronic devices, which employ nonequilibrium photoexcited carriers to improve the performance.
\end{abstract}

\maketitle

\section {Introduction}
  
Dynamics of nonequilibrium carriers in semiconductors greatly impact the performance of optoelectronic and photovoltaic devices because of the high excess energy that they can transfer \cite{Book_Jagdeep_Shah,WURFEL199743}.
The phenomenon is especially interesting in low-dimensional structures \cite{Ridley_1991}, where carriers can be extracted well before thermalization and potentially be exploited in external circuits \cite{Science2010Ti02}.
This approach, however, requires full understanding of the excited carrier evolution on ultrafast time scales.

In the following the term “hot carriers” refers to electrons or holes that are out of equilibrium. The carriers are either described by a Fermi-Dirac distribution with a temperature hotter than those of the lattice or,
in the strong nonequilibrium state when the concept of temperature becomes meaningless, they are instead characterized by a certain excess energy.
The simplest way to generate such hot carriers in a semiconductor is the interband excitation of electron-hole pairs by light absorption. The electron or hole excess energy, counted from the respective band edges, is determined by the frequency of the absorbed photons. Since we consider a single-temperature model in this work, we will use the terms "hot carriers" and "photoexcited carriers" interchangeably.
    
A direct band gap facilitates the interband absorption in semiconductors because of the vanishing momentum mismatch in optical transitions between the valence band maximum (VBM) and CBM. Group II-VI and III-V semiconductors exhibit a direct band gap and are therefore excellent candidates to exploit interband optical transitions. One of the popular choices among this group is CdTe, being a direct-band gap semiconductor used in electro-optic modulators and nuclear detectors \cite{CdTe_review}. Its most prevalent commercial application is in photovoltaics as a component of cadmium sulfide (CdS)-CdTe junctions \cite{CdS-CdTe-solar-cell}. State-of-the-art CdS-CdTe solar cells have reached impressive 21\% 
photovoltaic efficiency \cite{cdte_solarefficiency}, defined as the ratio of the electrical power output density to the incident light power input density.

\begin{figure}[!tb] \centering{}\includegraphics[width=1.0\columnwidth]{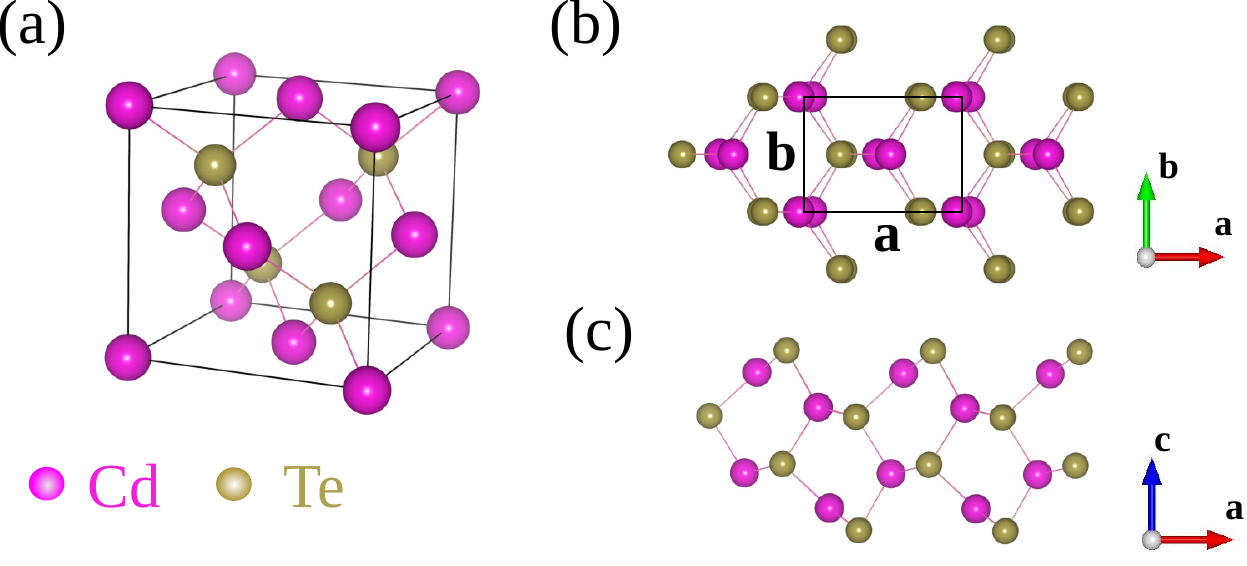}
    \caption{Energy-optimized crystal structure for (a) bulk CdTe in the zinc blende phase and monolayer CdTe cleaved along the [110] facet in (b) top and (c) side views. The lattice vectors $\mathbf{a},\mathbf{b},\mathbf{c}$ are indicated for each of the monolayer structures.} 
    \label{fig:CdTe-structure}
\end{figure}
    
Recent progress in nanofabrication technologies has made it possible to reduce the thickness of CdTe crystals to just a few layers \cite{Preparation_cdte_cdse_cds} and even to roll the layers up into nanoscrolls \cite{few-layer-CdTe-scrolls}. Atomically thin layers of other members of the group II-VI semiconductors have also been synthesized recently \cite{Preparation_cdte_cdse_cds, CdSe_nanosheet}. In particular, a four-layer-thick zinc selenide (ZnSe) slab cleaved along the [110] facet of the bulk crystal has been shown to be stable \cite{Znse_singlelayer}. Note that bulk CdTe features the same zinc blende crystal structure as ZnSe, see Fig.~\ref{fig:CdTe-structure}(a). Keeping in mind this tremendous progress, we anticipate fabrication of monolayer CdTe in the near future, because first principles studies suggest that this material is dynamically stable \cite{CdTe_simulation1, CdTe_simulation2, CdTe_simulation3}. In this work we address what happens with the ultrafast dynamics of photoexcited carriers in CdTe films, once their thickness is reduced down to the fundamental limit of a monolayer, as shown in Fig.~\ref{fig:CdTe-structure}(b).

Such extremely thin CdTe films belong to the class of atomically thin or 2D materials such as graphene \cite{Graphene_discovery}, transition metal dichalcogenides (TMDCs) \cite{MoS2_directgap_experiment_PRL}, 2D oxides \cite{oxide_discovery}, germanane \cite{Germanane_discovery}, and phosphorene \cite{Phosphorene_discovery}. 2D materials show properties drastically different from their bulk counterparts due to quantum confinement. For example, TMDCs demonstrate a transition from indirect to direct gap when reduced from the bulk to a monolayer, resulting in strong excitonic photoluminescence peaks \cite {MoS2_absorption, MoS2_directgap_experiment_PRL}. Interestingly, the opposite behavior of a transition from direct to indirect band gap can be observed in metal monochalcogenides, leading to the vanishing of excitonic absorption in stacks that are less than eight layers thick \cite{Budweg2019}. The peculiar optical properties make 2D materials promising for photodetection and other photonic applications \cite{TMDC_properties, Graphene_photonics, graphene_solarcell, Bernardi_am1.5, Mos2_solarcell, Linardy2020}.

To understand the microscopic details of hot-carrier dynamics in CdTe, both theoretical and experimental efforts are necessary. There are several pump-probe studies on the electron-spin relaxation in bulk CdTe, highlighting the role of photoexcited carrier density and excitation energy \cite{Spindynamics1, Spindynamics2, Spindynamics3}. In the pump-probe studies the carriers mainly thermalize via emission of optical phonons and interaction with defects \cite{Cdte_defects, Spindynamics1, Spindynamics3}. Although surface phonons in the group II-VI semiconductors have also been investigated \cite{Surfacephonons}, their role in photocarrier thermalization is still unknown. In perfect crystals the dominant mechanisms involved in hot-carrier scattering are electron-electron and electron-phonon interactions \cite{Ridley_1991}. It is well established by means of \textit{ab initio} methods that the electron-phonon scattering plays a major role near the band edges \cite{Neglect_e_e_interaction_TMDC, Bernardi_silicon, Bernardi_GaAs}. We therefore focus on the electron-phonon scattering in this paper and proceed similar to a previous work \cite{Yadav2019}, where we studied the photocarrier thermalization bottleneck in graphene. In what follows we combine DFT with many-body perturbation theory to calculate band- and momentum-dependent electron-phonon scattering times over a range of temperatures. Furthermore we use the Boltzmann equation in the RTA to compute the thermalization time of photoexcited carriers with a given excess energy. Our approach highlights the effects of electron-phonon scattering, providing access to the microscopic details involved in photocarrier thermalization processes. We show that the thermalization process is qualitatively different, depending upon whether CdTe is bulk or 2D. The respective thermalization times differ by orders of magnitude and demonstrate a distinct behavior upon excess energy and temperature changes. 

The paper is organized as follows. In Sec.~\ref{sec:compuational_details}, we discuss the computational methods used to calculate electronic and phononic properties, electron-phonon scattering rates, and photocarrier thermalization times. In Sec.~\ref{sec:results}, we explain the results obtained for bulk and monolayer CdTe. Finally, we conclude with a summary and an outlook in Sec.~\ref{sec:conclusions}.
    
\section{Computational details} \label{sec:compuational_details}

\subsection{\textit{Ab initio} approach to electron-phonon scattering times}\label{subsec:method-e-ph-structure}
    
To calculate electronic and phononic structure information, we use ground-state DFT within the local density approximation (LDA) considering spin-orbit coupling and rely on the implementation in \textsc{Quantum Espresso} \citep{QUANTUMESPRESSO_PW}. We employ plane wave (PW) basis sets with a kinetic energy cutoff of 70~Ry, a charge density cutoff of 280~Ry, and Troullier-Martins pseudopotentials including a relativistic core correction that assume six valence electrons for both Cd and Te \cite{TM_pseudopotential}. The dimensions of the unit cell and the atoms inside it are relaxed with the help of the Broyden-Fletcher-Goldfarb-Shanno algorithm for both bulk and monolayer CdTe until the net force on atoms is less than $10^{-6}$~Ry/a.u.\ and total energy changes are below $10^{-8}$~Ry. We separate CdTe monolayers in the out-of-plane direction through vacuum with a thickness of 18~\textup{\AA} to avoid spurious interlayer interactions. The electron density is calculated on a $\Gamma$-centered $\mathbf{k}$ grid in the BZ containing $12\times12\times12$ points for bulk and $9\times12\times1$ points for monolayer CdTe. The phonon band structure is computed within density functional perturbation theory (DFPT) \citep{DFPT_RevModPhys} on a $\Gamma$-centered $\mathbf{q}$ grid of size $6\times6\times6$ for bulk and $9\times12\times1$ for monolayer CdTe.
    
Having determined electronic and phononic band structures, we evaluate the electronic self-energy $\Sigma_{n\mathbf{k}}(T)$ in the lowest order of the electron-phonon interaction for a given electron band $n$, a wave vector $\mathbf{k}$, and the temperature $T$ using the \textsc{EPW} code \citep{EPW_package_PONCE2016, EP_wannier}. The self-energy is given as
\begin{widetext}
\begin{equation}
    \Sigma_{n\mathbf{k}}(T) = \sum_{m,p}\int_{\text{BZ}}\frac{d^3q}{\Omega_{\text{BZ}}}\vert
    g_{mn,p}(\mathbf{k},\mathbf{q})\vert^{2}\Bigg[\frac{N_{p\mathbf{q}}(T)+f^{(0)}_{m\mathbf{k+q}}(T)}{\en-(\varepsilon_{m\mathbf{k+q}}-\varepsilon_{\text{F}})+\hbar\omega_{p\mathbf{q}}+\text{i}\eta}\\
     +\frac{N_{p\mathbf{q}}(T)+1-f^{(0)}_{m\mathbf{k+q}}(T)}{\varepsilon_{n\mathbf{k}}-(\varepsilon_{m\mathbf{k+q}}-\varepsilon_{\text{F}})-\hbar\omega_{p\mathbf{q}}+\text{i}\eta}\Bigg].\label{eq:self-energy}
     \end{equation}
\end{widetext}
Here, $\hbar\omega_{p\mathbf{q}}$ is the phonon energy for the branch $p$ with the wave vector $\mathbf{q}$, $\varepsilon_{\text{F}}$ is the Fermi energy set to the middle of the gap for intrinsic semiconductors, $f^{(0)}_{n\mathbf{k}}(T)$ and $N_{p\mathbf{q}}(T)$ are the Fermi-Dirac and Bose-Einstein distributions for electrons and phonons, respectively, $\Omega_{\text{BZ}}$ is the volume of the BZ, and $\eta=20~\text{meV}$ is a small broadening parameter. The electron-phonon interaction matrix elements are a key ingredient in Eq.~(\ref{eq:self-energy}), and they are given by \citep{EPW_package_PONCE2016}  
\begin{equation}
  g_{mn,p}(\mathbf{k,q})=\frac{1}{\sqrt{2\omega_{p\mathbf{q}}}}\left<m\mathbf{k+q}|\partial_{p\mathbf{q}}V|n\mathbf{k}\right>,
\end{equation}
where $\partial_{p\mathbf{q}}V$ characterizes changes of the Kohn-Sham potential for displacements of nuclei through the phonon mode $p\mathbf{q}$. In other words, the matrix elements $g_{mn,p}(\mathbf{k,q})$ describe the electron scattering between the initial state $|n\textbf{k}\rangle$ and the final state $|m\textbf{k}+\textbf{q}\rangle$ mediated by the phonon $p\mathbf{q}$. The first term in the square brackets of Eq.~(\ref{eq:self-energy}) describes phonon absorption by an electron, the second one does so for phonon emission.
To compute $\Sigma_{n\mathbf{k}}(T)$, we interpolate electronic wavefunctions, eigenenergies and dynamical matrices on finer $\mathbf{k}$ and $\mathbf{q}$ grids using Wannier function (WF) techniques, as employed in the EPW code \cite{Wannier90, EP_wannier, EPW_package_PONCE2016,Wannier_orbitals}.
We find $\mathbf{k}$ and $\mathbf{q}$ grid sizes of $45\times45\times45$ for bulk and $300\times300\times1$ for the monolayer to be sufficient to converge the sum in Eq.~(\ref{eq:self-energy}).

The electron-phonon scattering time is determined as
\begin{equation}
  \tau_{n\mathbf{k}}(T)=\frac{\hbar}{2\text{Im}[\Sigma_{n\textbf{k}}(T)]}.
  \label{eq:tau-nk}
\end{equation}
Conversely, $\text{Im}[\Sigma_{n\textbf{k}}(T)]$ is proportional to the electron-phonon scattering rate $1/\tau_{n\mathbf{k}}(T)$ and is seen to be composed of contributions originating from phonon absorption and from phonon emission in Eq.~(\ref{eq:self-energy}).

We assume electron-phonon couplings to be small enough so that electronic bandstructures and eigenstates are basically unchanged as compared to their static ground-state values. Moreover lattice anharmonicity is disregarded for simplicity \citep{RevModPhys,EPW_package_PONCE2016}. Anharmonicity may be responsible for slow hot carrier cooling in halide perovskites, see Refs.~\cite{ReviewPerovskites2019,Perspectives2019} for recent reviews, and it can also be crucial in 2D crystals \cite{2DanharmonicPeters} because of flexural (out-of-plane) phonon modes, which are easy to excite.
The problem is well known for suspended graphene (see for instance Ref.~\cite{anharmonic-graphene-2020}, which provides an excellent introduction to this topic). However, technologically relevant samples are typically not suspended but placed on a substrate or even encapsulated. This stabilizes the 2D crystal by restricting the out-of-plane lattice motion. We therefore expect that anharmonic vibrational effects are substantially suppressed in real 2D samples and stick to the harmonic approximation for phonons as well as the lowest order expansion of the electron-phonon self-energy in terms of electron-phonon couplings.

\subsection{Time evolution of excited charge carriers}\label{subsec:method-time-evolution}
    
We simulate photoexcited-carrier dynamics in bulk and monolayer CdTe using the Boltzmann equation. The time evolution of the electronic occupation $f_{n\textbf{k}}(t,T)$ within the RTA is calculated separately for electrons and holes. It is given as
\begin{equation}
  \frac{df_{n\textbf{k}}(t,T)}{dt}=-\frac{f_{n\textbf{k}}(t,T)-f^{\mathrm{th}}_{n\textbf{k}}(T)}{\tau_{n\textbf{k}}(T)}   
  \label{eq:Boltzmann}
\end{equation}
with the solution
\begin{equation}
    f_{n\textbf{k}}(t,T)=f^{\mathrm{th}}_{n\textbf{k}}(T)+e^{-\frac{t}{\tau_{n\textbf{k}}(T)}}[f_{n\textbf{k}}(0,T)-f^{\mathrm{th}}_{n\textbf{k}}(T)].
    \label{eq:Boltzmann-solution}
\end{equation}
Here, $f^{\mathrm{th}}_{n\mathbf{k}}(T)$ is the Fermi-Dirac distribution at $t \to \infty$ with the quasi-Fermi level corresponding to the same number of photoexcited carriers as present at $t = 0$. 
Equation~(\ref{eq:Boltzmann-solution}) is only valid for low-intensity light excitations, leading to low densities of excited charge carriers, when the RTA applies \citep{Lundstrom_Book_lundstrom_2000}.
Optical excitations are furthermore assumed to be weak enough to avoid excessive sample heating. Hence, the lattice remains in equilibrium at temperature $T$, and the excited electrons relax to the Fermi-Dirac distribution at that lattice temperature.
Intermediate relaxation steps are omitted within the RTA, and the photoexcited carriers for a given $|n\mathbf{k}\rangle$ relax directly to thermal equilibrium with the rate $1/\tau_{n\mathbf{k}}(T)\propto \text{Im}[\Sigma_{n\textbf{k}}(T)]$. At higher intensities beyond the linear optical absorption regime RTA may fail, and a more sophisticated modeling is needed.

The initial photoexcited-carrier distribution $f_{n\mathbf{k}}(0,T)$ is modeled as the sum of the equilibrium Fermi-Dirac distribution $f^{(0)}_{n\mathbf{k}}(T)$ at the temperature $T$ and a nonequilibrium Gaussian centered at the excess energy 
$+\zeta$ ($-\zeta$) counted from the band edge for electrons (holes) as 
\begin{eqnarray}
     f_{n\textbf{k}}(0,T) & = & f^{(0)}_{n\textbf{k}}(T)-\lambda_{\text{h}}e^{\frac{(\varepsilon_{n\textbf{k}}+\zeta+\Delta/2)^{2}}{2\sigma^{2}}},\quad \varepsilon_{n\textbf{k}}<\varepsilon_{\text{F}}, \nonumber\\
     f_{n\textbf{k}}(0,T) & = & f^{(0)}_{n\textbf{k}}(T)+\lambda_{\text{e}}e^{\frac{(\varepsilon_{n\textbf{k}}-\zeta-\Delta/2)^{2}}{2\sigma^{2}}},\quad \varepsilon_{n\textbf{k}}\geq\varepsilon_{\text{F}}.\nonumber\\
     \label{eq:initial-distribution}
\end{eqnarray} 
Here, $\Delta$ is the electronic band gap and $\sigma = 8.47$~meV is a small energy broadening. We take the different DOS in the conduction and valence bands into account by adjusting $\lambda_{\text{e}}$ and $\lambda_{\text{h}}$ to ensure identical concentrations of photoexcited electrons and holes at any $\zeta$, separately for each material. For this purpose we define the two parameters via the energy-, time- and temperature-dependent population
\begin{equation}
    P(E,t,T)=\sum_{n\mathbf{k}}\delta(E-\varepsilon_{n\mathbf{k}})\times
    \begin{cases}
    [1-f_{n\mathbf{k}}(t,T)], E<\varepsilon_{\text{F}},\\
    f_{n\mathbf{k}}(t,T), E\geq\varepsilon_{\text{F}}.
    \end{cases}
    \label{eq:population}
\end{equation}
as
\begin{equation}
    \begin{split}
       \lambda_{\text{e}} = \frac{n_\text{e}  \alpha}{\int_{\varepsilon_{\text{F}}}^{\varepsilon_\text{max}}P(E,t = 0,T) dE}, \\
        \lambda_{\text{h}} = \frac{n_\text{h} \alpha}{\int_{-\varepsilon_\text{max}}^{\varepsilon_{\text{F}}}P(E,t = 0,T) dE}, 
    \end{split}
\end{equation}
where $n_\text{e}$ and $n_\text{h}$ are the photoexcited electron and hole densities, respectively, which we choose to be equal ($n_\text{e}=n_\text{h}$), 
and $\alpha$ is either the volume of the unit cell for bulk CdTe or the in-plane area of the unit cell for monolayer CdTe. In the expressions we integrate the population with a sufficiently large $\varepsilon_\text{max}=3$~eV symmetrically around the Fermi energy $\varepsilon_\text{F}=0$ (see also Fig.~\ref{fig:bandstructure-e-ph}).

Finally, we define the thermalization time $\tau_{\mathrm{th}}$, which we synonymously refer to as relaxation time, as
\begin{equation}
    \frac{P(\zeta,\tau_{\mathrm{th}},T)}{P(\zeta,0,T)}=\frac{1}{e}.
    \label{eq:tau_th}
\end{equation}
In our numerical calculations, we approximate the delta function in Eq.~(\ref{eq:population}) by a narrow Gaussian with a width of 20~meV. We focus on excitations sufficiently far above the quasi-Fermi levels such that the finite width does not affect $\tau_\mathrm{th}$.

\section {Results}\label{sec:results}
\subsection {Structural, electronic and phononic properties } 

\begin{figure}[!tb] \centering{}\includegraphics[width=1.0\columnwidth]{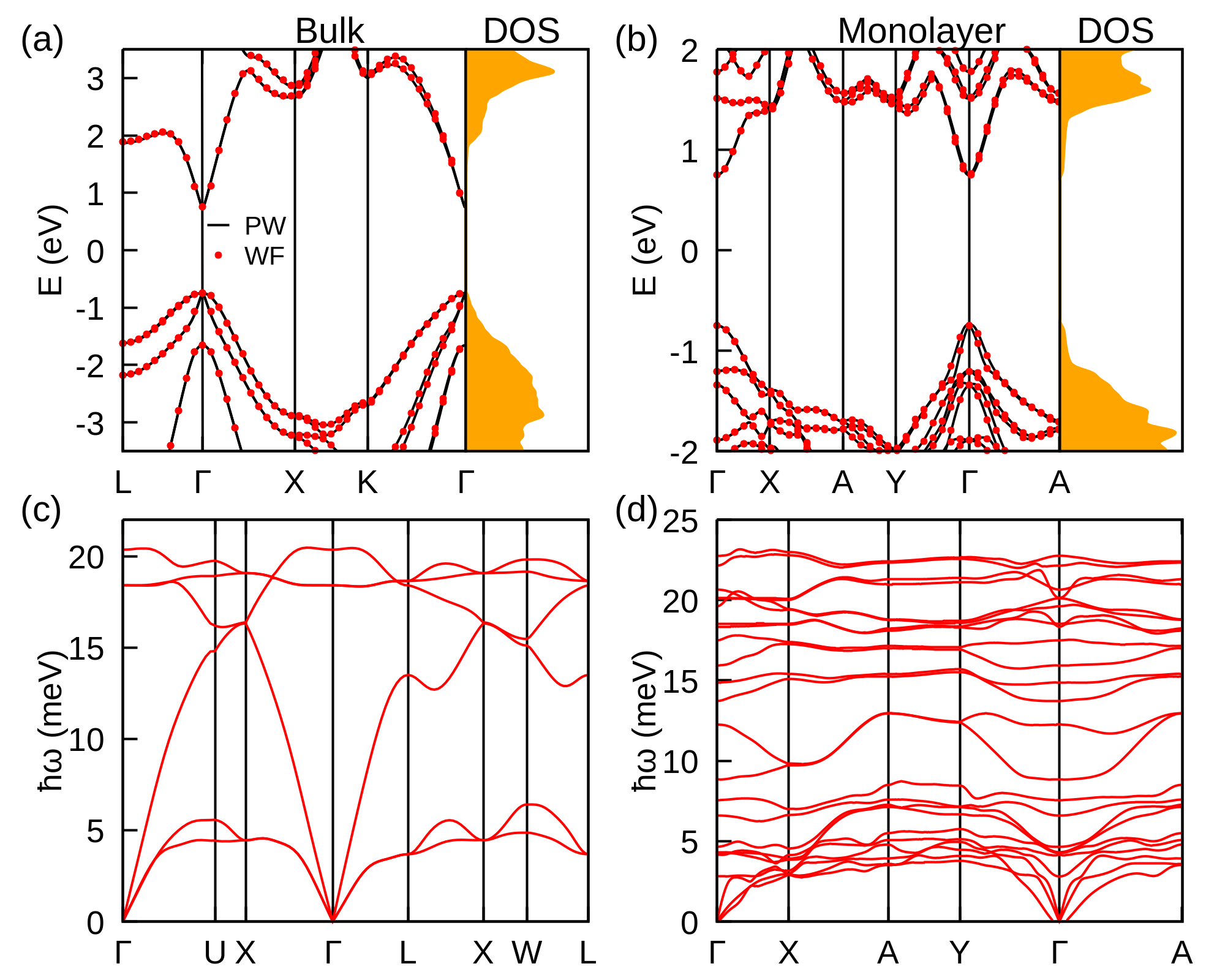}
    \caption{Electronic band structure calculated using plane wave (PW) and Wannier function (WF) basis sets for (a) bulk and (b) monolayer CdTe.
    The corresponding DOS is shown in arbitrary units to the right of the band structure. Phononic band structure for (c) bulk and (d) monolayer CdTe.}
     \label{fig:bandstructure-e-ph}
\end{figure}

Bulk CdTe adopts the zinc blende structure with a relaxed lattice constant of $|\textbf{a}| = 6.42$~\textup{\AA}, see Fig.~\ref{fig:CdTe-structure}(a), and shows Cd-Te bonds of mixed covalent-ionic character with lengths of $2.87$~\textup{\AA}, in good agreement with experimental data \cite{CdTe_bulk_latticeconstant}. All the facets except [110] exhibit a polar surface and are therefore expected to be unstable \cite{ZnSe_110}. This is the reason, why we cleave CdTe along the [110] facet to obtain a stable monolayer. As shown in Fig.~\ref{fig:CdTe-structure}(b) and \ref{fig:CdTe-structure}(c), the monolayer unit cell consists of eight atoms: four Cd, four Te, and each Cd atom is bonded to three Te atoms and vice versa. Due to the energy optimization of the unit cell, exterior Cd atoms are pulled into the layer, resulting in a rippled surface. We find lattice parameters of monolayer CdTe to be $|\textbf{a}|$ = 6.20~\textup{\AA}, $|\textbf{b}|$ = 4.48~\textup{\AA}, and Cd-Te bond lengths to vary in the interval of 2.72 to 2.83~\textup{\AA}, in agreement with a previous DFT study \cite{Phonon_monolayercdte}.  

Fig.~\ref{fig:bandstructure-e-ph}(a) and \ref{fig:bandstructure-e-ph}(b) shows the electronic band structures for bulk and monolayer CdTe along with the corresponding DOS. Both materials exhibit a direct band gap with the VBM and CBM located at the center of the BZ, i.e.\ at the $\Gamma$ point. As visible in Fig.~\ref{fig:bandstructure-e-ph}(a), the bulk valence band at the $\Gamma$ point is fourfold degenerate, and the spin-orbit split-off band is shifted down from the VBM by about 0.8~eV. The VBM of the monolayer is twofold degenerate, but the degeneracy is quickly lifted away from the $\Gamma$ point, see  Fig.~\ref{fig:bandstructure-e-ph}(b). We construct WFs from the PW basis, and the excellent match between WF-interpolated and PW-derived band structures demonstrates the quality of the WFs. DFT in the LDA is not able to predict the band gap values reliably, and more advanced methods are required for this purpose like DFT with hybrid exchange-correlation functionals or the GW approximation. We compute DFT band gap values for bulk and monolayer CdTe of 0.18 and 1.2 eV, respectively. To correct the band gaps, we perform a rigid shift of occupied and unoccupied electronic bands to match with the experimental value of 1.5~eV for bulk and with calculations using the Heyd–Scuseria–Ernzerhof (HSE) functional of 2.13~eV for monolayer CdTe \cite{Bulk_cdte_theory_experimentgap, Phonon_monolayercdte}. While the shift modifies the electronic gap sizes visible in Figs.~\ref{fig:bandstructure-e-ph}(a) and \ref{fig:bandstructure-e-ph}(b) and \ref{fig:ImSigma}, the temporal dynamics studied in Figs.~\ref{fig:P-time-evolution} and \ref{fig:thermalization-time} are not affected by this choice. Since the electronic band gap is much larger than maximum phonon frequencies, see Fig.~\ref{fig:bandstructure-e-ph}, electrons and holes thermalize separately in our model.
    
The phonon spectrum of bulk CdTe is well studied, and it shows a longitudinal-optical-transverse-optical (LO-TO) phonon splitting of around 2~meV at the $\Gamma$ point. Our result in Fig.~\ref{fig:bandstructure-e-ph}(c) matches well with the phonon band structure reported previously \cite{Phonon_bulk}. Monolayer CdTe, featuring eight atoms in the unit cell, gives rise to three acoustical and 21 optical modes with the highest longitudinal-optical mode energy of 23~meV reached between the $\Gamma$ and X points, see  Fig.~\ref{fig:bandstructure-e-ph}(d). This agrees with the DFT results of Ref.~\cite{Phonon_monolayercdte}. Positive frequencies throughout the BZ ensure the dynamical stability of monolayer CdTe. Imaginary frequencies around the $\Gamma$ point in monolayer CdTe are due to numerical problems in describing long wavelength phonons, and their magnitude is smaller than 0.11~meV. These modes are discarded in subsequent electron-phonon calculations with the EPW code \cite{EPW_package_PONCE2016} at the particular wavevectors, where they show imaginary frequencies.

\subsection{Self-energy and thermalization time}

\begin{figure}[!tb] \centering{}\includegraphics[width=1.0\columnwidth]{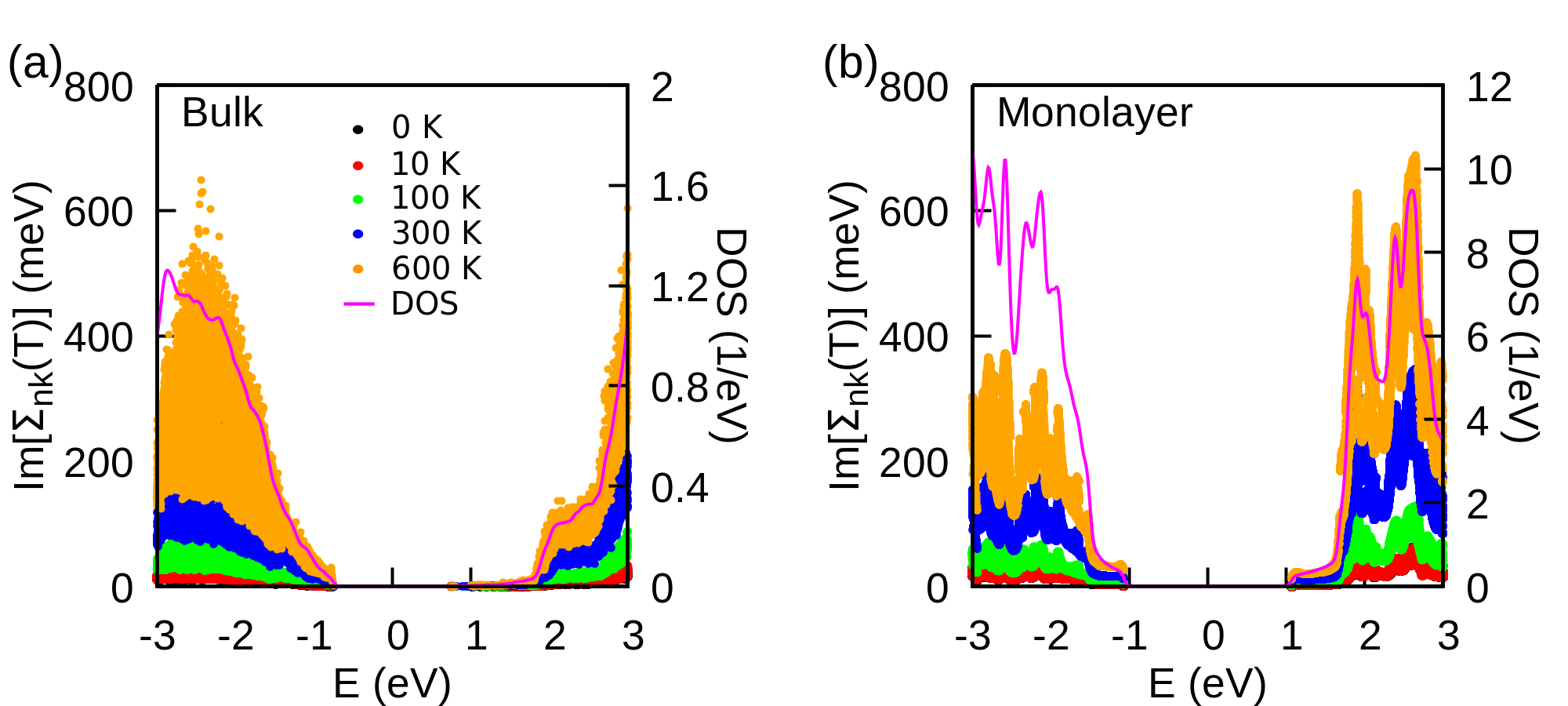}
    \caption{Imaginary part of the electron-phonon self-energy in Eq.~(\ref{eq:self-energy}) as a function of energy at different temperatures plotted together with the electronic DOS for (a) bulk and (b) monolayer CdTe. Panels (c) and (d) show the same data as panels (a) (b) on the logarithmic scale, with the DOS being omitted.}
     \label{fig:ImSigma}
\end{figure}

    Fig.~\ref{fig:ImSigma}(a) and \ref{fig:ImSigma}(c) shows the imaginary part of the self-energy Im[$\Sigma_{n\mathbf{k}}(T)$] for bulk CdTe as a function of energy at different temperatures on linear and logarithmic scales. 
    The imaginary part of the electronic self-energy exhibits small values from 0.75~eV up to 1.75~eV because of the twofold spin degeneracy in the conduction band, precluding interband scattering. At higher energy Im[$\Sigma_{n\mathbf{k}}(T)$] shows a sudden increase, reaching a plateau just above 2~eV that is followed by a second step at 2.7~eV. These abrupt changes around 1.75~eV and 2.7~eV take place when the CBM at the L and X valley, respectively, enter the energy window. The energy dependence of Im[$\Sigma_{n\mathbf{k}}(T)$] follows the electronic DOS, because the latter determines the available phase space for electron-phonon scattering events. In contrast to the degenerate conduction band, the presence of multiple valence bands allows for both inter- and intraband transitions. As a consequence, Im[$\Sigma_{n\mathbf{k}}(T)$] increases smoothly with energy for holes. For electrons and holes, the electron-phonon scattering increases in  efficiency at higher temperatures because the number of phonons increases, facilitating phonon absorption by both carrier types. 
    
    Fig.~\ref{fig:ImSigma}(b) and \ref{fig:ImSigma}(d) show Im[$\Sigma_{n\mathbf{k}}(T)$] for monolayer CdTe on linear and logarithmic scales. It keeps a constant value from $1.07$~eV up to 1.5~eV and then abruptly increases due to the local minimum in the conduction band around the Y point. In the case of holes, Im[$\Sigma_{n\mathbf{k}}(T)$] remains constant from $-1.07$~eV down to $-1.2$~eV and then also increases due to the second VBM at the $\Gamma$ point. The band structure becomes increasingly complex at higher excitation energies away from the band edges, and multiple extrema in the electronic band structure give rise to several peaks in Im[$\Sigma_{n\mathbf{k}}(T)$] for electrons and holes.  
    
\begin{figure}[!tb] \centering{}\includegraphics[width=1.0\columnwidth]{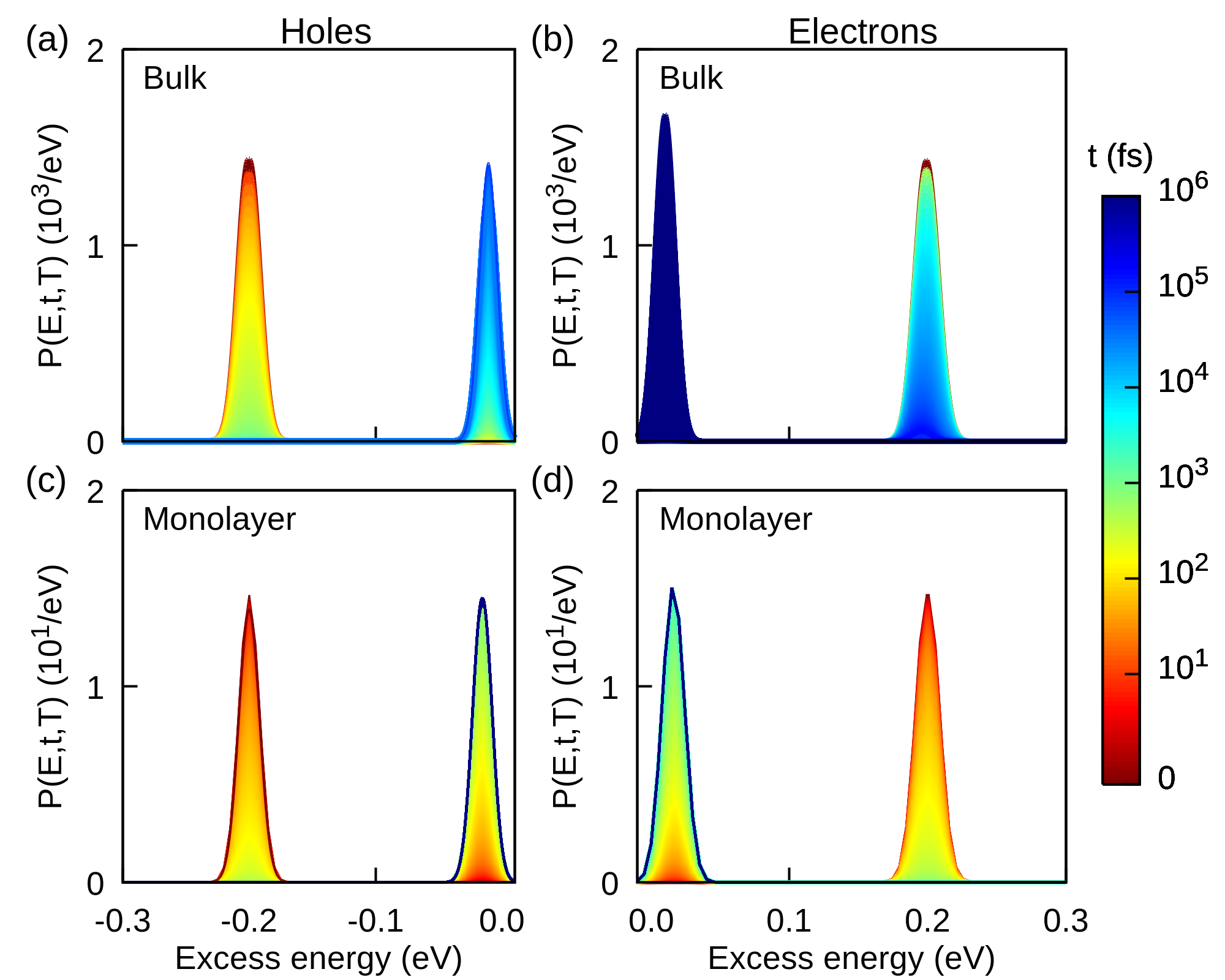}
    \caption{Temporal evolution of hole (left column) and electron (right column) populations, calculated separately using the Boltzmann equation in the RTA for (a),(b) bulk and (c),(d) monolayer CdTe at $T=10$~K. The carrier excess energy $\zeta$, counted from the respective band edges, is 0.2~eV. Initial and final populations correspond to photocarrier densities of $4.7 \times 10^{17}$~$\mathrm{cm}^{-3}$ for bulk and of $4.7 \times 10^{11}$~$\mathrm{cm}^{-2}$ for monolayer CdTe. The color scale indicates the time evolution (red for early, blue for late times). Note that the Gaussian shape of the thermalized population is a consequence of our choice to represent the delta function in Eq.~(\ref{eq:population}) by a Gaussian.} 
    \label{fig:P-time-evolution}
\end{figure}

    To analyze the photocarrier thermalization time, we set the reference level for the excess energy of hot holes and electrons to the valence and conduction band edges, respectively [see also Eq.~(\ref{eq:initial-distribution})]. Figure~\ref{fig:P-time-evolution} shows the evolution of the hot-carrier population, starting from the initial one comprising $4.7 \times 10^{17}$~cm$^{-3}$ and $4.7 \times 10^{11}$~cm$^{-2}$ hot carriers at an excess energy of 0.2~eV for bulk and monolayer CdTe, respectively. The thermalized population is obtained from the Fermi-Dirac distribution, with the quasi-Fermi level set to a value that reproduces the same carrier concentration as initially present. This assumption is valid as long as the photocarrier thermalization is much faster than the electron-hole recombination. Recombination times for carriers in bulk CdTe are reported to vary from 670 to 60~ns at concentrations from $10^{16}$ to $10^{17}$~cm$^{-3}$ \cite{Bulk_recombination}. 
    Figures~\ref{fig:P-time-evolution}(a) and \ref{fig:P-time-evolution}(b) show the hot carrier populations for bulk CdTe. The spin-degenerate conduction band limits the electron-phonon scattering, slowing down electron thermalization substantially, as illustrated by the blue color. In contrast, the holes thermalize much faster because of the broken degeneracy away from the VBM, allowing for both inter- and intraband scattering. Fig.~\ref{fig:P-time-evolution}(c) and \ref{fig:P-time-evolution}(d) demonstrate that electrons and holes thermalize approximately at the same rate in monolayer CdTe, but the rate is much higher than in the bulk case. The reason becomes clear from Fig.~\ref{fig:bandstructure-e-ph}(c) and \ref{fig:bandstructure-e-ph}(d): Monolayer CdTe possesses a rich phonon spectrum that provides much more scattering channels than its bulk counterpart.
\begin{figure}[!tb] \centering{}\includegraphics[width=1.0\columnwidth]{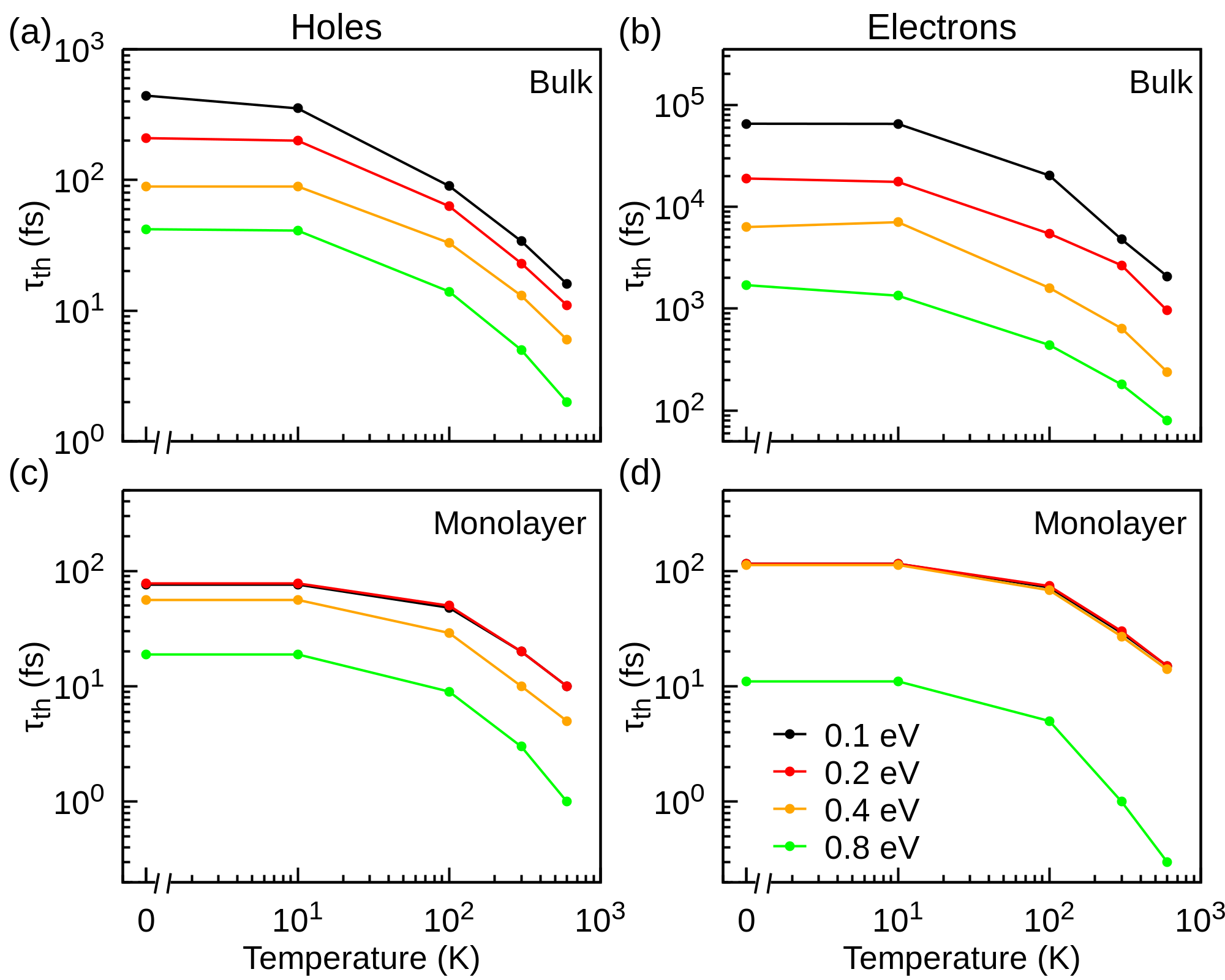}
    \caption{Thermalization time for holes (left column) and electrons (right column) as a function of temperature over a range of excess energies for (a),(b) bulk and (c),(d) monolayer CdTe. Solid lines show the thermalization time, when the FI is only included in the phonon dispersion ("partial FI"), while dashed lines for the bulk consider the effect of the FI also in the electron-phonon couplings ("full FI").}
    \label{fig:thermalization-time}
\end{figure}

Figure~\ref{fig:thermalization-time} summarizes our key findings regarding the thermalization time $\tau_\mathrm{th}$, defined in Eq.~(\ref{eq:tau_th}). We study $\tau_\mathrm{th}$ as a function of temperature over a range of excess energies $|\zeta|$ = 0.1, 0.2, 0.4, 0.8~eV. In the case of bulk CdTe, holes thermalize approximately two orders of magnitude faster than electrons at the same excess energy, see Figs.~\ref{fig:thermalization-time}(a) and \ref{fig:thermalization-time}(b). In the case of monolayer CdTe, the hole and electron thermalization times are of the same order of magnitude. But interestingly there is either very weak or almost no change of $\tau_\mathrm{th}$ for electrons up to an excess energy of about 0.5~eV due to the degenerate conduction band and the constant DOS in this region. Thermalization times generally shorten with increasing temperature, and the reduction accelerates at temperatures above 100~K. Figures.~\ref{fig:thermalization-time}(c) and \ref{fig:thermalization-time}(d) suggests extremely fast thermalization ($\tau_\mathrm{th}<20$ fs) in monolayer CdTe at high temperatures about 600~K. Nearly identical thermalization times for electrons and holes at low excess energies arise from a comparable DOS around the conduction and valence band edges.

It is instructive to compare our computational findings regarding thermalization times with experimental pump-probe spectroscopy data. We note first of all that it is the excitation energy ($\hbar\omega$) that is controlled experimentally rather than electron and hole excess energies. Electron and hole excess energies may differ ($\zeta_\mathrm{e} \neq \zeta_\mathrm{h}$) for a given photon energy, if electron and hole effective masses are not equal ($m_\mathrm{e} \neq m_\mathrm{h}$). The excess energies then relate to the excitation energy as 
\begin{equation}
    \zeta_\mathrm{e,h} = \frac{m_\mathrm{h,e}}{m_\mathrm{e}+m_\mathrm{h}}\left(\hbar\omega - \Delta\right).\label{eq:excess-energy}
\end{equation}
The formula states for instance that if holes are much heavier than electrons, electrons are excited much farther away from the band edge than holes. Our results still remain applicable, if electrons and holes thermalize independently within their respective bands.
    
In bulk CdTe, valence and conduction bands are asymmetric, and holes are much heavier than electrons. From the electronic dispersion in Fig.~\ref{fig:bandstructure-e-ph} we extract hole and electron effective masses at $\Gamma$ along the $\Gamma$-X direction of $0.80m_0$ and $0.09m_0$ \cite{PhysRev.129.2466}, respectively, where $m_0$ is the free electron mass. Most of the excitation energy is thus transferred to electrons. The asymmetry is, however, reduced in monolayer CdTe, which exhibits a more uniform distribution of excess energy between electrons and holes. We calculate effective masses of carriers at the $\Gamma$ point along the $\Gamma$-X direction of $0.71m_0$ and $0.39m_0$ for holes and electrons, respectively.
    
Since CdTe consists of different atoms, polar interactions need to be examined for a comparison to experiment, in particular the interaction of electrons with LO phonons. In the results studied so far, the Fr\"ohlich interaction (FI) has been taken into account in the phonon dispersion only. This yields the relaxation times shown with solid lines in Fig.~\ref{fig:thermalization-time}. If we take the FI into account also in the electron-phonon couplings within \textsc{EPW} \citep{EPW_package_PONCE2016}, we obtain the results shown with dashed lines. We determined these results only for the bulk, because the calculations for the monolayer turned out to be computationally too demanding due to the larger number of atoms in the unit cell. The results show quantitatively that thermalization times are reduced through full consideration of the FI, but qualitatively the trend of a $\tau_\text{th}$ that decreases with increasing $T$ and $\zeta$ remains the same. We attribute this behavior to the electronic DOS that obviously does not depend on FI. Hence, we expect the qualitative predictions regarding the thermalization time in 2D CdTe to be valid even though FI is disregarded in electron-phonon coupling matrix elements.

Measurements are not yet available for monolayer CdTe, but thermalization times for the bulk can be estimated from optical experiments \cite{KimelPRB,Leitenstorfer_cdte}. Kimel {\em et al.}\ have identified two relaxation processes with respective decay times of about 100~fs and 1~ps \cite{KimelPRB} at room temperature by means of the Kerr effect. They have used an excitation energy of 1.63~eV, which is above the electronic band gap of 1.52~eV. Due to the largely different electron and hole masses in bulk CdTe [Fig.~\ref{fig:bandstructure-e-ph}(a)] and the relation in Eq.~(\ref{eq:excess-energy}), this translates to an electron excess energy of about 0.1~eV that corresponds to the black curve in Fig.~\ref{fig:thermalization-time}(b), and a hole excess energy that is around one order of magnitude lower. For the electron dynamics, which is detected in the experiments, our modeling thus yields a relaxation time of some 500~fs [see Fig.~\ref{fig:thermalization-time}(b)] if we include FI at room temperature. The longer time scale of the order of ps is attributed to spin-relaxation processes \cite{KimelPRB}, which is consistent with other experiments \cite{Spindynamics1, Spindynamics2, Spindynamics3}. The fact that we overestimate the shorter time scale is expected, since experimentally more scattering channels, such as electron-impurity or electron-electron scattering, should be present than assumed in our simulations of pure electron-phonon scattering. Another experiment has measured an electron relaxation time of about 70~fs at a low temperature of 20~K and at low excitation densities by means of differential transmission spectroscopy with the pump pulse centered at 1.65~eV \cite{Leitenstorfer_cdte}. Our calculated time scale of some 3~ps is consistently above the measured 70~fs because such a low temperature is freezing out the phonons.

\section{Conclusions and outlook}\label{sec:conclusions}

We have studied carrier thermalization in bulk and monolayer CdTe within a range of up to 0.8~eV away from the band edges. At low temperature, the photoexcited carriers in bulk CdTe thermalize on a 0.1--100 ps timescale, whereas in monolayer CdTe this process completes within 10-100 fs. Increased temperatures generally reduce the thermalization time. 
Our observations are consistent with the current understanding that reduced dimensionality alters the electron-phonon scattering rate \cite{Ridley_1991} and consequently the carrier thermalization.
The thermalization time also depends on the excitation energy and is in general substantially reduced, if the carriers are excited far away from the band edges. Our most unexpected observation is that the photoexcited electrons in monolayer CdTe comprise an exception from this rule: Their thermalization time does not depend on excess energy up to about 0.5~eV above the CBM. We attribute the behavior to a parabolic conduction band that forms in the monolayer and whose band minimum is well separated from the other conduction bands.

The model that yields the results quoted can be improved in various ways. First of all, the FI is only taken into account in the phonon dispersion. If it is considered in the electron-phonon couplings, we have shown that thermalization times are quantitatively reduced in the bulk, while qualitative trends remain unaffected. Unfortunately, such computationally demanding calculations on the whole BZ including relativistic spin-orbit coupling could not be accomplished here for the monolayer. Additional progress towards more realistic simulations may be made by using two-temperature approaches that simultaneously describe the thermalization in the coupled electronic and phononic subsystems \cite{2TM_metal_1, 2TM_metal_2, 2TM_PRL}, by going beyond the RTA \cite{perturbo, Jhalani2017}, or to quantum models \cite{kadanoff2014entropy,KB_equation}.

One of the promising strategies for increasing the amount of work done per absorbed photon is to speed up the photocarrier extraction into an external electrical circuit \cite{Science2010Ti02}. The carriers are then collected while they are still hot or even out of thermal equilibrium. The saved excess energy can then be utilized for increasing the photoresponse. 2D materials might offer unprecedented opportunities along this path: Their thickness is so small that the photoexcited carriers could be extracted well before they thermalize and reach ambient temperature by means of phonon emission. The mechanism has recently been studied in a graphene-based van der Waals heterostructure \cite{ma2016tuning}.
The hot photocarriers are filtered out by a potential barrier so that only the carriers with high enough excess energy can contribute to the photocurrent. However, the photocarrier thermalization time usually drops substantially with increasing excess energy, as one can see in Fig.~\ref{fig:thermalization-time}(a)-\ref{fig:thermalization-time} (c). This is also true for other popular 2D semiconductors, including MoS$_2$ or WSe$_2$ \cite{Yadav:PRR2020}. Hence, high excess energy and slow thermalization are two conflicting requirements which are difficult to satisfy at the same time. Our main result, shown in Fig.~\ref{fig:thermalization-time}(d), suggests that monolayer CdTe is a great exception: Electron thermalization does not significantly depend on excess energy. Clever utilization of this effect may help to circumvent the problem of hot photocarrier extraction.

\section{Acknowledgements}
We thank A. Leitenstorfer for inspiring discussions. D.Y.\ and F.P.\ acknowledge financial support from the Carl Zeiss Foundation as well as the Collaborative Research Center (SFB) 767 of the German Research Foundation (DFG) at University of Konstanz. M.T. acknowledges the Director's Senior Research Fellowship from the CA2DM at NUS (Singapore NRF Medium Sized Centre Programme R-723-000-001-281) and thanks the Okinawa Institute of Science and Technology for its hospitality. Part of the numerical modeling was performed using the computational resources of the bwHPC program, namely the bwUniCluster and the JUSTUS HPC facility.

\bibliography{cdte.bib}
\end{document}